\documentclass[aps,showpacs,twocolumn]{revtex4}
\usepackage[dvips]{graphics,graphicx}  
\begin{document}   
\title{Quantum Chaos in the Bose-Hubbard model}   
\author{Andrey R. Kolovsky$^{1,2}$ and Andreas Buchleitner$^1$} 
\affiliation{$^1$Max-Planck-Institut f\"ur Physik komplexer Systeme,  
             D-01187 Dresden} 
\affiliation{$^2$Kirensky Institute of Physics, Ru-660036 Krasnoyarsk.}  
\date{\today} 
   
\begin{abstract}  
We present a numerical study of the spectral properties 
of the 1D Bose-Hubbard model. Unlike the 1D Hubbard model for fermions, 
this system is found to be non-integrable, and
exhibits Wigner-Dyson spectral statistics under suitable conditions.
\end{abstract}   
\maketitle  
 
During the last two decades much attention has been paid to the 
spectral properties of quantum systems which exhibit chaotic 
dynamics in the classical limit. The intrinsic complexity of these systems  
(due to underlying chaos) leads to an  
irregular energy spectrum with universal 
statistical properties -- a phenomenon often referred
to as `quantum chaos' \cite{Gian91}. One of  
the recent developments in the field is related to `many-body 
quantum chaos' \cite{Izra00}. Here the systems of interest consist of
many interacting (identical) particles and usually have no 
classical counterpart.  Notwithstanding, numerical
studies of some simple models (like, for example, a 1D chain of 
interacting spins \cite{Mont93}) have shown
that the spectrum of a deterministic many-body system \cite{remark1}  
can also exhibit universal statistics. In the present paper
we numerically study the spectrum of the 1D Bose-Hubbard model  
\cite{Fish89}. Our particular interest in this system is due to  
its possible laboratory realization with  
cold atoms loaded into optical lattices \cite{Grei02}.
It is shown below that, unlike the 1D Hubbard model for fermions,  
the 1D Bose-Hubbard model is generally non-integrable. Moreover,  
in a certain parameter regime its spectrum obeys
the Wigner-Dyson statistics for Gaussian Orthogonal Ensembles and,  
hence, this system can be considered as a quantum chaotic system.
 
The Hamiltonian of the 1D Bose-Hubbard (BH) model reads  
\begin{equation}   
\widehat{H}= - 
\frac{J}{2}\left(\sum_l \hat{a}^\dag_{l+1}\hat{a}_l +h.c.\right)   
+\frac{W}{2}\sum_l\hat{n}_l(\hat{n_l}-1) \;.  
\label{1}   
\end{equation}  
Having in mind spinless atoms in a quasi 1D  
optical lattice, the creation, $\hat{a}^\dag_l$, or  
annihilation operator $\hat{a}_l$ in (\ref{1}) `creates' or  
`annihilates' an atom in the $l$th well of the optical potential, in a 
Wannier state $\psi_l(x)$. The parameter $J$ is the hopping matrix 
element, the parameter $W$ the on-site interaction energy 
for atoms sharing one and the same well. The Hilbert 
space of the system (\ref{1}) is spanned by Fock states 
given (in the coordinate representation) by symmetrised 
products of single-particle Wannier states. Assuming a 
finite size of the lattice, $l=1,\ldots,L$, and a finite number $N$  
of the atoms (which is a conserved quantity),  
the dimension of the Hilbert space is ${\cal N}=(N+L-1)!/N!(L-1)!$. 
  
It is worth noting that instead of the Wannier basis one can use a basis 
given by symmetrised products of Bloch waves
$\phi_\kappa(x)=(1/\sqrt{L})\sum_l \exp(i\kappa l)\psi_l(x)$,  
where $\kappa=2\pi k/L$ is the single-particle quasimomentum 
($\kappa=2\pi$ corresponds to the centre of the Brillouin zone).
In second quantisation, this change of the
basis amounts to the substitution 
$\hat{b}_k=(1/\sqrt{L})\sum_l \exp(i2\pi kl/L)\hat{a}_l$. 
Then the Hamiltonian (\ref{1}) takes the form 
(up to an additive term $WN/2$) 
\begin{equation} 
\label{1b} 
\widehat{H}=-J\sum_k \cos\left(\frac{2\pi k}{L}\right) 
\hat{b}_k^\dag\hat{b}_k 
\end{equation} 
\begin{displaymath} 
+\frac{W}{2L}\sum_{k_1,k_2,k_3,k_4} 
\hat{b}_{k_1}^\dag\hat{b}_{k_2}\hat{b}_{k_3}^\dag\hat{b}_{k_4} 
\tilde{\delta}(k_1-k_2+k_3-k_4) \;, 
\end{displaymath} 
where $\tilde{\delta}(k)=1$ if $k$ is a multiple of $L$, and 
$\tilde{\delta}(k)=0$ otherwise (conservation of total  
quasimomentum). 
   
We proceed with our analysis of the spectrum of the BH model.   
In the limits $J=0$ (no hopping) and $W=0$ (non-interacting bosons), 
the BH model is completely integrable, with eigenstates 
given by symmetrised products of the single-particle Wannier 
states $|n_1,\ldots,n_l,\ldots,n_L\rangle$, or by symmetrised 
products of Bloch states  $|n_1,\ldots,n_k,\ldots,n_L\rangle$,  
respectively. The corresponding eigenvalues are 
\begin{equation}   
E_{\mu}=W\mu \;,\quad \mu=\frac{1}{2}\sum_{l=1}^L n_l(n_l-1)  
\;,\quad \sum_{l=1}^L n_l=N \;, 
\label{2}   
\end{equation}  
for $J=0$, and 
\begin{equation}   
E_{\nu}=-J\nu \;,\quad  
\nu=\sum_{k=1}^{L} \cos\left(\frac{2\pi k}{L}\right) n_k  
\;,\quad \sum_{k=1}^L n_k=N \;, 
\label{3}   
\end{equation} 
for $W=0$. Note that in both cases the majority of the energy levels  
are multiply degenerate, i.e. different combinations of 
integers $n_l$ ($n_k$) may result in the same 
values of $\mu$ ($\nu$).  
\begin{figure}[t!]    
\center    
\includegraphics[width=8.5cm, clip]{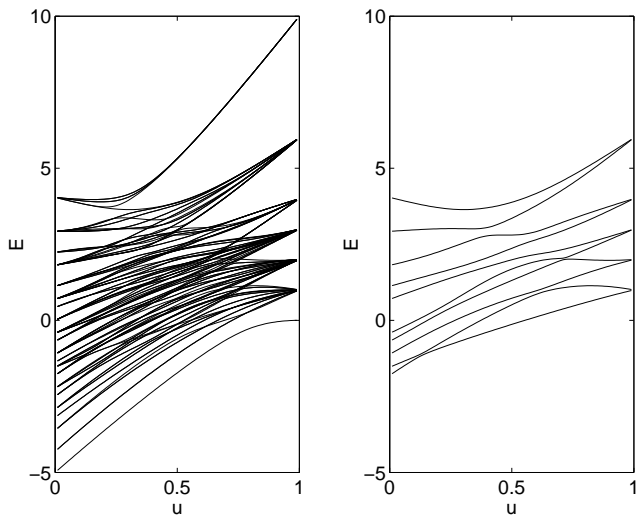}    
\caption{Energy level dynamics of the BH model, as a functions of  
the parameter $u\equiv W$, $J=1-u$, for $L=N=5$ 
and periodic boundary conditions. The left panel shows
the complete spectrum, the right panel singles out the
`odd $\kappa=2\pi$' symmetry component.} 
\label{fig2}    
\end{figure} 
 
Now we demonstrate the non-integrability of the BH model for $J,W\ne0$. 
The left panel in Fig.~\ref{fig2} shows its spectrum for
finite size $L=N=5$, as a function of the parameter $u=W$, with $J=1-u$. 
The first step of the analysis is to account for the trivial symmetry
of the spectrum, due to the translational invariance of 
the Hamiltonian. This implies that the Hamiltonian (\ref{1},\ref{1b})
factorises as
\begin{equation} 
\label{5} 
H=\bigoplus_{k=1}^L H^{(\kappa)} \;, 
\end{equation} 
where $\kappa=2\pi k/L$ is now the {\em total} quasimomentum. 
Thus, the spectrum in Fig.~\ref{fig2}(a) is a   
superposition of $L$ independent spectra. Next we account for the
`odd-even' symmetry of the $\kappa=\pi$ and $\kappa=2\pi$ subspaces. 
The levels of the odd $\kappa=2\pi$ symmetry are shown in the right 
panel of Fig.~\ref{fig2}. No crossings between levels are left 
(on the scale of the figure, some tiny avoided crossings are not 
resolved), and, thus, no further decomposition
of the spectrum is possible. This proves the non-integrability  
of the system \cite{Neum29}. It is interesting to note in this context
that the 1D Hubbard model for fermions is integrable for arbitrary $u$.
In the latter case, in addition to the trivial ($u$-independent)  
symmetries (i.e., quasimomentum and odd-even symmetry), the system
has a number of non-trivial ($u$-dependent) integrals \cite{Shas86,Yuzb02}.
Therefore, the spectrum of the 1D Hubbard model can be 
decomposed further until only one (possibly degenerate) level is left.

The above non-integrability of our system (manifest in the
avoided crossings in the right panel of Fig.~\ref{fig2})  
does not yet guarantee that the spectrum 
of the BH model has the universal statistical properties 
associated with quantum chaos. Loosely speaking, the latter requires
a delocalisation of the eigenfunctions of the system in any generic basis.
To quantify such delocalisation, we consider the Shannon entropy
\begin{equation} 
\label{6a} 
S(u)=\big\langle -\frac{1}{\log {\cal N}} \sum_{j=1}^{\cal N} 
|c_j|^2 \log(|c_j|^2)\big\rangle \;, 
\end{equation} 
where the $c_j$ are the expansion coefficients of an arbitrary 
eigenstate of the system in a given basis, and the 
angular brackets denote an average over all eigenstates. 
It is reasonable to choose as a basis the `$\nu$-basis',
when our starting point is $u=0$ (no interaction), or the
`$\mu$-basis', when starting at $u=1$ (no hopping).
First let us clarify the notion of $\nu$- and $\mu$-basis. 
As mentioned above, and as apparent
from Fig.~\ref{fig2}, the spectrum of the  
system at $u=1$ ($J=0$) consists of a number of degenerate levels 
with energies given in Eq.~(\ref{2}). An arbitrarily small 
$J$ removes this degeneracy, and the levels split into a 
$\mu$-band with a width proportional to $J$. Moreover, 
there is a vicinity of $J=0$ where the eigenfunctions of 
the system do not depend on $J$. These states form the $\mu$-basis. 
More formally, the $\mu$-basis is given by the eigenstates 
of the Hamiltonian (\ref{1}) where the matrix elements 
between Fock states with different $\mu$ are set (by hand)  
to zero. Analogously, the $\nu$-basis is given by the eigenstates 
of the Hamiltonian (\ref{1b}) where the matrix elements 
between states with different $\nu$ are set to zero. 
\begin{figure}[b!]    
\center    
\includegraphics[width=8.5cm, clip]{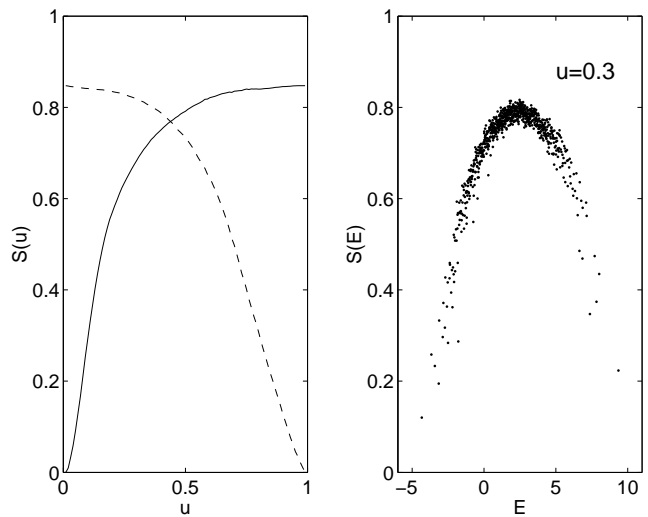}    
\caption{Left panel: Mean entropy (\ref{6a}) in the $\nu$-basis 
(solid line) and in the $\mu$-basis (dashed line),
as a function  of the interaction parameter $u$ ($N=L=8$). Right panel:
Entropy (\ref{6b}) of the individual eigenstates (dots), for $u=0.3$.}
\label{fig3}    
\end{figure}
 
The entropy (\ref{6a}) calculated in the $\nu$-basis 
(solid line) and in the $\mu$-basis (dashed line), respectively,  
is depicted in the left panel of Fig.~\ref{fig3}, for a filling
factor ${\bar n}=N/L=1$. There is a relatively wide interval
of $u$ where the functions $S^{(\nu)}(u)$ and $S^{(\mu)}(u)$  
{\em simultaneously} take large values. Thus, the 
eigenfunctions of the system are essentially delocalised in 
either of the two basis sets, and one may expect universal 
spectral statistics. Indeed, a  statistical analysis of the  
spectrum confirms this expectation.
\begin{figure}[t]
\center
\includegraphics[width=8.5cm, clip]{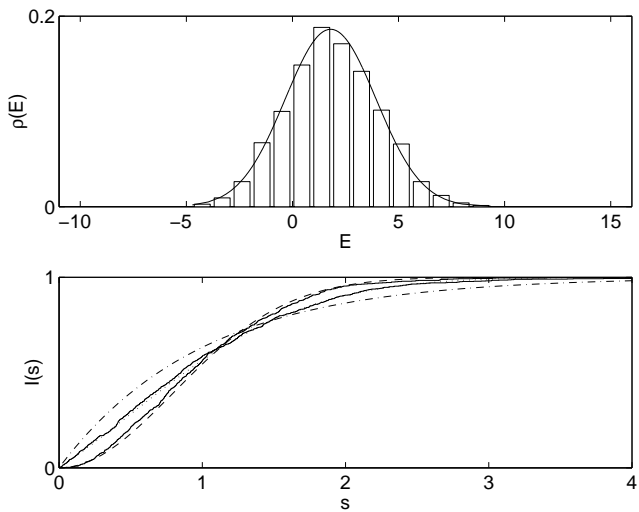}    
\caption{Density of states (upper panel) and integrated level spacing
distributions (lower panel) for $L=N=8$ and $u=0.3$.  
The solid line in the upper panel is the best Gaussian fit to 
the numerically obtained density of states. 
The dash-dotted, dashed and dotted lines in the lower panel represent
the random matrix predictions (\ref{8}-\ref{10}) for Poissonian, Gaussian 
orthogonal (GOE), and two superimposed, independent GOE ensembles, 
respectively. The solid lines are obtained from the numerical data for 
$\kappa=2\pi$ (which reproduce the expectation for superimposed GOE spectra), 
and for $\kappa=2\pi/8$ (which faithfully obey the GOE prediction).}
\label{fig4}    
\end{figure}
The histogram in the upper panel of Fig.~\ref{fig4} shows the 
distribution of eigenenergies $E$ for $L=N=8$ and $u=0.3$. 
The numerically obtained density of states
is well approximated by a Gaussian distribution, 
$\rho(E)\sim\exp[-(E-\bar{E})^2/\sigma^2]$, 
where $\bar{E}$ and $\sigma$ are fitting parameters \cite{remark2}.
Using $\bar{E}$ and $\sigma$ from the best fit we unfold the spectrum 
and extract the nearest-neighbour level statistics for $\kappa=2\pi$ and
$\kappa=2\pi/L$. The result is presented 
in the lower panel of Fig.~\ref{fig4}, by the integrated 
level spacing distribution $I(s)$, $s=\Delta E\rho(E)$ (solid line),  
in comparison to the Random Matrix Theory predictions for the Poissonian
(dash-dotted),  
\begin{equation} 
\label{8} 
I(s)=1-\exp(-s)\, , \; 
\end{equation} 
and the Gaussian orthogonal ensemble (dashed),
\begin{equation} 
\label{9} 
I(s)=1-\exp\left(-\frac{\pi s^2}{4}\right)\, , \; 
\end{equation} 
as well as for the case resulting from a superposition of  
two independent GOE spectra (dotted line),
\begin{equation} 
\label{10} 
I(s)=1+\exp\left(-\frac{\pi s^2}{16}\right) 
\left[-1+\rm{erf}\left(\frac{\sqrt{\pi}s}{4}\right)\right] \;. 
\end{equation} 
The numerical data faithfully follow the distributions
(\ref{9},\ref{10}) for quasimomentum $\kappa=2\pi/L$
and $\kappa=2\pi$, respectively, where the odd and the even part of the
spectrum are superimposed in the latter case.
   
We now address the conditions for quantum chaos to prevail in the BH model.
Obviously, the above criterion based on the mean entropy (\ref{6a}) 
provides only a rough estimate on the parameter region where one
may expect an irregular spectrum. Moreover, besides 
the interaction parameter $u$ and the filling factor $\bar{n}$, 
also the energy $E$ should be considered as a relevant parameter.  
As a an illustration of this statement, the right panel in Fig.~\ref{fig3}  
shows the entropy of the individual eigenstates, 
\begin{equation} 
\label{6b} 
S(E)=\min[S^{(\nu)}(E),S^{(\mu)}(E)] \;, 
\end{equation} 
which are labeled here by the individual level's energy $E$.  
Obviously, there is a nonvanishing fraction of  
localised eigenstates associated with the lower and upper parts of the  
spectrum. Hence, one has to distinguish regular and chaotic spectral 
components \cite{Stas95}. In the presently
considered case ($u=0.3$, $\bar{n}=1$) the regular component 
is negligible (less than 1/10 of the total number  
of levels) and, hence, the level spacing distribution 
follows the universal Wigner-Dyson law. However, when the ratio $J/W$
approaches either one of the integrable limits, the regular component
gradually increases at the expense of the irregular one,  
what causes a deviation from 
the universal distribution. The same holds true if we decrease the 
value of the filling factor. In this latter case the chaotic 
component (typically located in the central part of the spectrum)  
is small even for $J/W\simeq 1$. Further studies  
are needed for a precise mapping of the chaotic region  
of the BH model in the parameter space spanned by $u$, $\bar{n}$, and $E$. 
\begin{figure}[t]    
\center    
\includegraphics[width=8.5cm, clip]{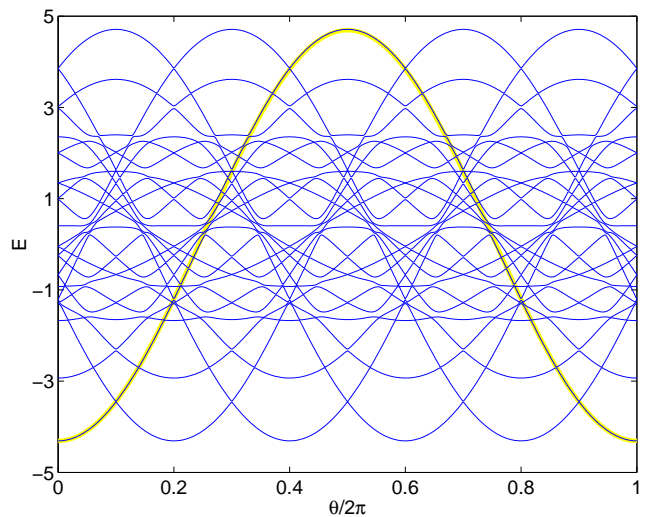}
\caption{Level dynamics of the modified BH model (\ref{11}) 
for $u=0.1$, $\kappa=2\pi$, and $L=N=5$. The grey shaded energy level
is the diabatic approximation for the ground state evolution,
which neglects the interaction induced avoided crossings.}   
\label{fig5}
\end{figure} 
 
Let us now discuss the physical manifestations of the irregular spectrum
of the BH model, i.e. its relevance for ongoing experiments 
with cold atoms in optical lattices \cite{Grei02,Daha96,Roat04}.
Obviously, to probe the irregular part of the spectrum  
one has to excite the system. This can be done, for example, by applying  
a static force, i.e., adding a term $F\sum_l l \hat{n}_l$ to the Hamiltonian 
(\ref{1}). Note that the translational symmetry of the
system, broken by this term, can be recovered by a suitable 
gauge transformation \cite{pre}. This leads to a time-dependent  
Hamiltonian of the form 
\begin{equation}   
\widehat{H}(t)= -
\frac{J}{2}\left(e^{-iFt}\sum_l \hat{a}^\dag_{l+1}\hat{a}_l +h.c.\right)
  +\frac{W}{2}\sum_l\hat{n}_l(\hat{n_l}-1) \;,
\label{11}   
\end{equation} 
and a parametric dependence of the eigenvalues of the Hamiltonian (\ref{11})
on $\theta=Ft$, as illustrated in Fig.~\ref{fig5} (for $L=N=5$, $\kappa=2\pi$,
as in Fig.~\ref{fig2}, and $u=0.1$).
Let us assume for a moment that there are no atom-atom
interactions and, thus, all avoided crossings in Fig.~\ref{fig5}
are true crossings. Then, in the course of time, the static
force will drive the system along the continuation of the ground state
(the grey shaded energy level in Fig.~\ref{fig5}), representing nothing
else than bona fide Bloch oscillations,
where the system comes back to its initial state after 
one Bloch period $T_B=2\pi/F$. It is clear, however, that 
for $u\ne 0$ the system generally does not come back to
the initial state, because of Landau-Zener transitions at the
avoided crossings encountered during one Bloch cycle \cite{remark3}.
In other words, after the Bloch cycle is completed, some excited states
will be populated as a consequence of (a)diabatic transitions along 
the diabatic continuation of the ground state. (The branching
ratio between diabatic and adiabatic transitions is obviously determined 
by the time derivative of $\theta$, i.e., by the field amplitude $F$.)
In the case of strong static forcing, $F\gg J$, this excitation process
is reversible, leading to {\em quasiperiodic} Bloch
oscillations \cite{prl1}. In contrast, for moderate forcing (and
chaotic instantaneous spectrum \cite{remark4}) the excitation
is irreversible, leading to decoherence of the one-particle density 
matrix and, as a consequence, to decay of the Bloch oscillations \cite{prl2}.
Since Bloch oscillations of the atoms are easily measured in laboratory 
experiments \cite{Daha96,Roat04}, their {\em irreversible decay} provides
a direct indication of chaos in the BH system \cite{Roat04}.

In conclusion, we have shown that the 1D Bose-Hubbard model is
a generic quantum chaotic system with characteristic spectral irregularity.
This is in stark contrast to the 1D Hubbard model for fermions, 
which is known to be an integrable system with regular spectrum.
It is also worth noting that the {\em many-site} character of the BH model 
is crucial for the appearance of quantum chaos -- the two site BH model, 
which is often used to study Bose Einstein condensates
of cold atoms in double-well potentials \cite{doublewell}, 
has a regular spectrum.

 

\begin{thebibliography}{10}
\bibitem{remark2}
In the present case of moderate interaction strength, 
the parameters $\sigma$ and
$\bar{E}$ scale with the number of atoms $N$ as $\sigma\sim J\sqrt{N}$ and
$\bar{E}\sim W\bar{n}^2N$. Let us also note that for strong
interaction the density of states is asymmetric, and
the approximation of $\rho(E)$ by a Gaussian is no more valid.
\bibitem{remark3}
Note that all levels in Fig.~\ref{fig4} anticross, at finite $u$, except 
for exact crossings between even and odd symmetry states at 
$\theta_j=(\pi/L)j$, $j=0,\ldots,2L-1$.
\bibitem{remark4}
Statistical analysis of the instantaneous spectrum (not shown here)
suggests the following simple rule: If the spectrum
at $\theta=0$ exhibits universal random matrix statistics,
it equally does so for $\theta\ne0$.
For the presently considered case $\kappa=2\pi$ and $L=N$ this
means that the spectrum reproduces Eq.~(\ref{10}) for $\theta=\theta_j$
\protect\cite{remark3}, interpolates between (\ref{10})
and (\ref{9}) in the vicinity of these symmetry points, 
and obeys Eq.~(\ref{9}) for other values of $\theta$.
\end{thebibliography}
\end{document}